\documentclass[aip, pop, amsfonts, amssymb, amsmath ,preprint,subscriptadress]{revtex4-1}
\usepackage{amssymb}		 
\usepackage{amsmath, amsthm, amssymb,amsfonts}
\usepackage[english]{babel}
\usepackage{color, graphicx}
\usepackage{caption}
\usepackage{comment}
\usepackage{hyperref}

\def \ov{\over}
\def \beq{\begin{equation}}
\def \eeq{\end{equation}}
\def \lb{\label}

\def \pd{\partial}
\def\~#1{\widetilde #1}

\def \sy {symmetry}
\def \sys {symmetries}
\def \so {solution}
\def \eq{equation}

\def\la{\lambda}

\def \q{\quad} 
\def\={\, =\, }
\def\ni{\noindent}
\def\sk{\smallskip}

\begin{document}

\title{MHD equilibria with incompressible flows: symmetry approach}
\author{G. Cicogna}\affiliation{Dipartimento di Fisica ``E.Fermi'' dell'Universit\`a di Pisa
and INFN, Sez. di Pisa, Largo B. Pontecorvo 3,  I-56127, Pisa, Italy} 
\author{F. Pegoraro}\email{pegoraro@df.unipi.it} \affiliation{Dipartimento di Fisica ``E.Fermi'' dell'Universit\`a di Pisa, Largo B. Pontecorvo 3, 56127 Pisa, Italy}

\begin{abstract}
We identify and discuss a  family of  azimuthally symmetric, incompressible, magnetohydrodynamic   plasma  equilibria with poloidal and toroidal flows  
in terms of solutions  of the Generalized Grad Shafranov  (GGS) equation.
These solutions are  derived by exploiting   the  incompressibility assumption,  in order to rewrite the GGS equation in terms of a different dependent variable,  and the continuous Lie symmetry properties of the resulting equation  and in particular a special type of  ``weak''  symmetries.
\end{abstract}


\maketitle

\section{Introduction}
There is a rich literature concerning the transformation properties and the symmetry analysis of  the  general form  of the equations that  describe  the magnetohydrodynamic  (MHD) equilibrium of  a plasma with flows. See e.g., Ref.\onlinecite{GK}, where transformation properties of a generalized Grad--Shafranov (GS) equation  
are analyzed for cases with and without spatial symmetry in a framework that makes it possible to recover both the magnetostatic and the ``stationary hydromechanic''  limits. 
In Ref.\onlinecite{PN}, as an application of the method,  solutions of the stationary three-dimensional MHD equations with field-aligned incompressible flows are calculated.
Other applications of the  transformation properties as a useful method for finding explicit solutions for the general   MHD equilibrium equations 
 starting from known solutions can be found in Refs.\onlinecite{B1,B2,B3}, whereas in Ref.\onlinecite{Ch} the equivalence  is shown of this method  with the classical  approach to infinitesimal Lie point symmetries.

The symmetries of interest   in this analysis are described by continuous Lie groups of transformations and we refer e.g., to the books \onlinecite{Ol,CRC,BA}  for all details about this method and for some of its typical applications in different physical problems.
It can be remarked that the presence of a symmetry  has twofold applications: i)  it makes it possible to obtain new solutions starting from a known one, ii) it makes it possible to look for solutions which are left invariant by the symmetry transformation (this may be obtained considering suitable  {\it reduced} equations).

The analysis of the  symmetry properties of the specific GS equation that describes the case of stationary, azimuthally symmetric MHD   plasma  equilibria  without  flows  (static equilibria) was developed in previous papers, see e.g. Refs.\onlinecite{Sig,WH,CND,PoP10} and   Ref.\onlinecite{FOG} for the  Euler equations with swirl.
In  Ref.\onlinecite{Sig} an analysis of the symmetry  properties of a   class of partial differential  equations of relevance for plasma physics, including the GS equation as a  special case, has been performed. In particular, in Ref.\onlinecite{PoP10}   the notion of  ``weaker'' symmetries\ (conditional  symmetries  and similar) was introduced for the GS equation; the presence of such symmetries  made it possible  to identify  for the first time some additional classes of solutions to the GS  equation that describe D-shaped toroidal plasma equilibria.  The present paper is devoted to the investigation of a generalized form  of  the GS  equation
that describes two-dimensional azimuthally symmetric   plasma equilibria in the presence of poloidal and toroidal plasma flows, see Refs.\onlinecite{lovelace,goedbloed,AP8}, but we will restrict our  study to incompressible flows. The rationale behind the incompressibility assumption
stems from the fact that it has been known in the literature, Refs.\onlinecite{Morr,TT,TT1,TT2, AMP12}, that  in this limit  the equation that takes the role of the GS
in the presence of plasma flows, Eq.(\ref{GSL})  in Sec.\ref{GGS}, can be transformed  into a much simpler  equation of the   GS-type.
This is made possible by a redefinition of the dependent variable (the magnetic flux function $\psi$) and can be applied as long as  the poloidal  flow is sub-alfv\`enic\cite{ext}.
This method allows us to generalize to a plasma with flows a class of solutions identified  in Ref.\onlinecite{PoP10} using this symmetry approach.

\bigskip 

The present paper is organized as follows. \,  In Sec.\ref{GGS}  we  introduce the Generalized Grad Shafranov  (GGS) equation which includes the effect of poloidal and toroidal flows, 
take the incompressible plasma limit, define the new dependent variable $\chi = \chi(\psi)$ and obtain a modified GS equation with an additional term that has an $r^4$ dependence of the cylindrical radius $r$.
\\ In Sec.\ref{WCS}  we apply the concept of weak conditional symmetry to the GS 
equation.  First we recover the D-shaped solution of the GS equation described in  Ref.\onlinecite{PoP10} as a special case of a wider class of solutions. The common geometrical structure in the poloidal $r$-$z$ plane   of this class of  solutions is discussed, and shown to arise directly from its symmetry properties.  On the other hand, determining the specific spatial shape of  these solutions involves the (numerical) integration of a nonlinear second order ODE whose coefficients depend of the physical parameters of the specific equilibrium configuration. A qualitative classification of the different types of solutions of this ODE is given. \\ In Sec.\ref{SyGGS} we extend this analysis  to the  GGS equation and describe a novel class of equilibria solutions with flows. These solutions share the same common geometrical structure of the solutions without flows because the symmetry properties are the same. Specific examples of  equilibria with flows are  then shown,  including solutions  that  exhibit  three  magnetic axes along  the toroidal direction. These latter solutions   correspond  to configurations made of  three separate sets of tori, each set nested around one of the magnetic axes. By imposing appropriate   boundary conditions corresponding to $\psi $=${\bar \psi}$=$ const$ surfaces, each of these sets can be considered as an independent equilibrium configuration.  \\In the Conclusions  we  note that in the case where poloidal  flows are present  the reconstruction from the solutions of  the GGS equation  of the explicit spatial  dependence of the plasma pressure,   of the poloidal and toroidal velocity  fields and of the toroidal field   from that of the  variable $\chi$ involves  the inversion of the integral relationship between $\psi $ and $\chi$ together with the solution of a set of nonlinear algebraic  equations.  This inversion is not required in the case of toroidal flows  and an explicit example of the effect of the toroidal flows  and their gradients on the equilibrium configuration is shown. \\ Finally, we indicate further  extensions of the solution procedure of the GS and of the GGS equations employed  in the present article. 

\section{Generalized Grad Shafranov equation}\label{GGS}

We consider an  axisymmetric Magnetohydrodynamic plasma equilibrium with poloidal ($v_p$) 
and azimuthal  ($v_ \varphi$) flows and 
 write  the  corresponding generalized Grad Shafranov equation (see Refs.\onlinecite 
{lovelace,goedbloed,AP8})
in the form
\begin{eqnarray}
\mathbf{\nabla\cdot}\left[  \left(
1-\frac{\mathcal{F}^{2}}{4\pi\rho}\right)
\frac{1}{r^{2}} \,{\mathbf{\nabla}\psi}\right]  +\frac{\mathcal{FF}^{^{\prime}}%
}{4\pi\rho} \frac{\left\vert  \mathbf{\nabla}\psi\right\vert
^{2}}{r^2} = -4\pi\rho\left(
\mathcal{J}^{\prime} - U_{S}\mathcal{I}^{\prime}  + rv_{ \varphi}\mathcal{G}^{\prime}\right)
\nonumber\\
\hspace{ 2.0 cm}
-\frac{1}{r^{2}}(\mathcal{H}+rv_{ \varphi}\mathcal{F})(\mathcal{H}^{\prime}+rv_{ \varphi}\mathcal{F}^{\prime}).
\label{GSL}%
\end{eqnarray}
Here the notation of Ref.\onlinecite{AP8} has been adopted with coordinates 
$\left(r, \varphi,z\right)$ where the azimuthal angle
$ \varphi$\ is an  ignorable coordinate. In Eq.(\ref{GSL}) a prime denotes differentiation 
with respect to the magnetic  flux function $\psi$, \, $\rho(r,z)$ is the plasma density and the magnetic field $\mathbf{B}$  is expressed  in terms of its  azimuthal and
poloidal components, $B_{ \varphi}$ and $\mathbf{B}_{p}$ through  the flux function $\psi(r,z)$ as 
\begin{equation}
\mathbf{B} = B_{ \varphi}\mathbf{\hat{ \varphi}}+\mathbf{\nabla}\psi\times
\mathbf{\nabla} \varphi \label{eq:Bdef}\,,
\end{equation} where $\mathbf{\hat{ \varphi}}=r\mathbf{\nabla} \varphi$ is
the unit vector in the azimuthal direction.  The flux functions in Eq.(\ref{GSL}) are defined as 
\begin{eqnarray}
\mathcal{F}  \left(  \psi\right)   =4\pi\rho  v_p /B_p   ,\label{eq:ffun1}\\
\mathcal{G}\left(  \psi\right)     = v_{ \varphi}/{r} \, - \, {\mathcal{F}B_{ \varphi}}/({4\pi \rho \, r}),
\label{eq:ffun2}  \\
\mathcal{H}\left(  \psi\right)     =rB_{ \varphi}-r\mathcal{F}v_{ \varphi}%
,\label{eq:ffun3}\\
\mathcal{I}\left(  \psi\right)     =S,\label{eq:ffun4}\\
\mathcal{J}\left(  \psi\right)     =U+\rho U_{\rho}+{v^{2}}/{2}-rv_{ \varphi
}\mathcal{G}. \label{eq:ffun5}%
\end{eqnarray}
 where $S$ is the plasma entropy, $U$ is the plasma internal energy ad $U_\rho$ its derivative at constant entropy.\\
 In the following we consider an  incompressible equilibrium with $\rho =  const.$ in which case  $\rho( \mathcal{J}^\prime - U_{S}\mathcal{I}^{\prime}) = ( p  + \rho  {v^{2}}/{2}-  \rho rv_{ \varphi
}\mathcal{G})^\prime$ and define the  Alfv\`enic Mach number $M(\psi) =  \mathcal{F}/(4 \pi \rho)^{1/2}$.\,
Inverting Eqs.(\ref{eq:ffun2}-\ref{eq:ffun3}) we obtain 
\begin{eqnarray}
r  B_{ \varphi}    = [ r^2 {\mathcal{F}}{\mathcal{G} }+ {\mathcal{H}}]/(1 - M^2), \label{BF}\\
 r v_{ \varphi}  =  [r^2  {\mathcal{G} }+ {\mathcal{H}}{\mathcal{F}}/(4\pi \rho )]/(1 - M^2).
\label{VF}
\end{eqnarray}
The r.h.s. of Eq.(\ref{GSL}) contains functions of $\psi$ and powers of $r$. By substituting for $ r v_{ \varphi}$ its expression from Eq.(\ref{VF}) and collecting powers of $r^2$, we can rewrite the  r.h.s. of Eq.(\ref{GSL})  as \\
\begin{equation}
-\frac{1}{2 r^2}\,\left[\frac{\mathcal{H}^2}{ 1-M^2} \right]^\prime
-\left[  
\left( \frac{ \mathcal{G} \mathcal{F}\mathcal{H}}{1-M^2}\right)^\prime \,+ 4\pi\rho\left(
\mathcal{J}^{\prime} - U_{S}\mathcal{I}^{\prime}\right)  \right]  -  r^2\left[\frac{ (\mathcal{G}^2 {\mathcal{F}^2}/2)^\prime }{ 1-M^2}  +4\pi \rho({\mathcal{G}^2}/2)^\prime  \right]. \label{rhs}\end{equation}

In the limit of zero plasma flow the $1/r^2$ term reduces to  the toroidal magnetic field contribution   to  the  standard GS  equation and the $r$-independent  term to the plasma pressure  contribution.
\,
The plasma poloidal  flow modifies  the operator on the l.h.s.  of Eq.(\ref{GSL}) as discussed in Refs.\onlinecite{lovelace,goedbloed}.  On the r.h.s.  the poloidal flow introduces the multiplication factors  depending on $(1-M^2)^{-1}$ and  both flows modify the flux functions in the $1/r^2$ in  the $r$-independent term. In, addition they introduce  a new term proportional to $r^2$. In the remaining part of this  article we will consider only sub-alfv\`enic poloidal flows so that  $M^2 <1$.

\subsection{A new dependent variable: $\chi = \chi(\psi)$}

The fact that  the alfv\`enic Mach number $M$    in a constant density equilibrium  is a flux function allows us to  bring back the operator on the l.h.s.   of Eq.(\ref{GSL}) to its standard GS form by redefining  the dependent variable $\psi$. 
Following Refs.\onlinecite{Morr,TT,TT1,TT2,AMP12} we define the new dependent variable $\chi = \chi[\psi(r,z)]$  as 
\begin{equation}\label{Morr}
\chi(\psi) = \int ^\psi [1 - M^2(\eta)]^{1/2} d \eta,
\end{equation} 
where $M^2(\eta) <1$. \, Then, we obtain 
\begin{equation}\label{tr} 
\mathbf{\nabla\cdot}\left[  \left(
1-\frac{\mathcal{F}^{2}}{4\pi\rho}\right)
\frac{1}{r^{2}} \,{\mathbf{\nabla}\psi}\right]  +\frac{\mathcal{FF}^{^{\prime}}%
}{4\pi\rho} \frac{\left\vert  \mathbf{\nabla}\psi\right\vert
^{2}}{r^2} =   (1-M^2)^{1/2} \mathbf{\nabla\cdot}\left[   \frac{1}{r^{2}}\,{\mathbf{\nabla}\chi}\right]  .
\end{equation}
Finally, re-expressing  the flux functions of $\psi$ on the r.h.s. of Eq.(\ref{rhs}) as functions of $\chi$ through their dependence on $\psi(\chi)$ and using cylindrical coordinates, we obtain (see also
 Ref.\onlinecite{TH14,TH14-1})
\begin{equation}\label{ecce}
\frac{\partial^2  \,  {\chi} }{\partial \, r^2} - \frac{1 }{r} \frac{\partial \,  {\chi} }{\partial \, r}  + \frac{\partial^2  \,  {\chi} }{\partial \, z^2} = {\cal A}_0({\chi}) +  r^2 \,  {\cal A}_2({\chi}) +   r^4 \,  {\cal A}_4({\chi}) ,
\end{equation}
where 
\begin{equation}\label{ecce1}
 {\cal A}_0({\chi})  \equiv \frac {1}{2 [1-M^2]^{1/2}} 
 ,\left[\frac{\mathcal{H}^2}{1-M^2} \right]^\prime ,  \end{equation}
 \begin{equation}\label{ecce2}
{\cal A}_2({\chi})  \equiv \frac {1}{ [1-M^2]^{1/2}}  \left[  
\left( \frac{ \mathcal{G} \mathcal{F}\mathcal{H}}{1-M^2}\right)^\prime \,+ 4\pi\rho \left(
\mathcal{J}^{\prime} - U_{S}\mathcal{I}^{\prime} \right)  \right] ,
\end{equation}
\begin{equation} \label{ecce3}
 {\cal A}_4({\chi})    \equiv \frac {1}{ [1-M^2]^{1/2}} \left[\frac{ (\mathcal{G}^2 {\mathcal{F}^2}/2)^\prime }{ 1-M^2}  +4\pi \rho({\mathcal{G}^2}/2)^\prime \right]  .
  \end{equation}

\section{Symmetries, Conditional and   ``weak''  Conditional Symmetries.}\label{WCS}

\subsection{GS equation with no flow: ${\cal A}_4=0$}

Let us start by considering the ``standard'' GS \eq\ (i.e. with no flow so that in particular  {\it $\chi(\psi) = \psi$})
\beq\lb{GSst}    {\pd^2\psi\ov{\pd r^2}}-{1\ov r}{\pd\psi\ov{\pd r}}
+{\pd^2\psi\ov {\pd z^2}}\=r^2F(\psi)+G(\psi)  . \eeq 
The classification  presented in Ref.\onlinecite{PoP10} 
exhausts all the possible Lie point-sym\-me\-tries  admitted by this \eq .  The presence of these \sys\ is strictly related to precise choices of the functions $F(\psi)$ and $G(\psi)$, as fully discussed in  Ref.\onlinecite{PoP10}.
Other papers devoted to the \sy\ analysis of this \eq\ (or of particular cases of it) are e.g.  Refs.\onlinecite{Sig,WH,CND,FOG}.
In Ref.\onlinecite{Rum}, also an example of  {\it generalized} \sy\   is proposed.

As shown in Ref.\onlinecite{PoP10}, one can also  look for  weaker notions of symmetries, 
and in particular for the existence of {\it conditional} symmetries  Refs.\onlinecite{LW,FK,OR} (see also Refs.\onlinecite{CND,CL} { for other references}).
Let us recall that a conditional symmetry\ $Y$ generates a 
transformation which does {\it not} map \so s into \so s, but defines a 
$Y$-invariant variable with the property that the \so s of the { reduced equation obtained writing the initial one}  in terms of this variable,  are also \so s of the initial \eq . 

In Ref. \onlinecite{PoP10}, we have shown to be useful to introduce  even {\it weaker} type of conditional \sy\
(see Ref.\onlinecite{CND,OR,CL} for this notion of \sy ); in particular, the choice
\beq\lb{wcs} Y\= z{\pd\ov{\pd r}}+ r{\pd\ov{\pd z}}\eeq
has revealed to be especially convenient (see Ref.\onlinecite{PoP10} for a discussion and other details). 
Writing the GS \eq\ in terms of the $Y$-invariant variable
\[s\=  r^2-z^2\]
one obtains the following \eq , where the variable $r$ still appears (playing here the role of a parameter, as happens in the case of weak conditional \sys )
\begin{equation} (8   r^2-4s)\psi_{ss}-2\psi_s\=r^2F(\psi)+G(\psi) \label{special}.
\end{equation}
This \eq\ can be naturally split into two ODE's involving only   the new variable $s$:
\beq\lb{due} 8\psi_{ss}\=F(\psi) \q\q{\rm and}\q\q\ 4 s  \psi_{ss}+2\psi_s\= -\,G(\psi)\ . \eeq
Clearly, these \eq s admit \so s only if there is a precise relationship 
between $F(\psi)$ and $G(\psi)$. A simultaneous solution  of these \eq s can be found of 
the form
\beq\lb{saq}\psi(s)\=(r^2-z^2)^{-q} \eeq
for any real $q\not=0$, with
\beq\lb{wsol}F(\psi)\=8q(q+1)\psi^{1+(2/q)}\q\q,\q\q G(\psi)\=-2q(2q+1)\,\psi^{1+(1/q)}\ .\eeq
Let us notice that with the choice (\ref{wsol}) for $F,\,G$, the GS \eq\ admits the Lie point 
scaling \sy
\beq\lb{sy1} X_1\=r\frac{\pd}{\pd r}+z\frac{\pd}{\pd z}-2\,q\,\psi \frac{\pd}{\pd\psi} \eeq
and that the above \so\ (\ref{saq}) is invariant under this \sy .

\subsection{A D-shaped equilibrium \so .}\label{D}

The choice  $q=-1/4$ in (\ref{saq}) is of special interest as it makes it possible to construct a new class of solutions using an exceptional symmetry of the GS equation. 
We consider the \so\cite{note}
\beq\lb{unq} \psi\=(r^2-z^2)^{1/4}\= s^{1/4}\eeq
which clearly holds only in the region $|z|\le r$, and solves the GS \eq\ with
\beq F(\psi)\=-(3/2)\psi^{-7}\q,\q  G(\psi)\=(1/4)\psi^{-3} \eeq
As shown in Ref.\onlinecite{PoP10}, with the above choice for $F,\,G$ and with $q=-1/4$, the GS \eq\ 
admits also the ``exceptional'' \sy\ (see also Ref.\onlinecite{FOG})
\beq\lb{exc}X_2\= 2rz{\pd\ov{\pd r}}+(z^2-r^2){\pd\ov{\pd 
z}}+z \psi {\pd\ov{\pd \psi}}\ .\eeq
The presence of this \sy\ 
implies that if $\psi(r,z)$ is a \so\ of the GS \eq , then also
\beq\lb{solex}
\~\psi(r,z)\=[C(r,z,\la)]^{1/2}\psi\big(\~r(r,z,\la),\~z(r,z,\la)\big)\eeq
where
\[C(r,z,\la)\=1+\la^2(r^2+z^2)+2\la z\]
\[\~r\=r\,[C(r,z,\la)]^{-1}\q\q\ {\rm and}\q\q\ 
\~z\=[z+\la(r^2+z^2)]\,[C(r,z,\la)]^{-1}\]
is a \so\ of the \eq\ for any value of the real parameter  $\la$. 
Taking  advantage from  this \sy , we can construct from the \so\ (\ref{unq})
a continuous family of \so s (see Ref.\onlinecite{PoP10}): we
obtain, for any   $\la$,
\beq\lb{Dsh}\psi(r,z)\=[ r^2-\big(z+\la  (r^2+z^2)\big)^2]^{1/4}\eeq
which holds in the interior of the two circles 
centered resp. in $r_0\!=1/(2\la)$, $z_0\!= -1/(2\la)$ and 
$r_0\!=-1/(2\la)$, $z_0\!=\!-1/(2\la)$, both of radius
$1/(\sqrt{2}\la)$, excluding  their intersection (it is
not restrictive to assume $\la>0$). 
Thanks to the obvious invariance of the GS \eq\ under translations
$z\to z + const.$, we can  shift this \so\ and choose $z_0=0$. 
\,
The resulting ``D-shaped'' \so s and configurations are shown and fully discussed in Ref.\onlinecite{PoP10}.
\sk\ni

\subsection{Additional ``exceptional symmetry'' solutions.}\label{AES}

It is possible to  generalize the D-shaped   solution described above  by  
taking  \beq F(\psi)\= a_2  \psi^{-7}\q,\q  G(\psi)\=a_0 \psi^{-3} \eeq 
where  $a_0$ and $a_2$  are  free real  coefficients and 
by setting 
\begin{equation}\label{Dgen0} {\psi}(r,z) = s ^{1/4}\,   \phi(y), \qquad y = r^2/s  \ge 1 .  \end{equation}
The variable   $y$ {is an invariant variable under the symmetry (\ref{sy1}) and } is a trigonometric function of the ``latitude''  angle $\theta$,\, ($1- 1/y =  z^2/r^2  = \tan^2{\theta}$) with $y = 1$  corresponding to the equatorial ($z = 0$) plane. 
For the sake of definiteness we  restrict ourselves to   positive  values of $a_0$  and conveniently  rescale the variable $\psi$ such that  for all cases  $a_0 = 1/4$.
Inserting  Eq. (\ref{Dgen0}) into  Eq.(\ref{GSst}) one obtains, thanks to the invariance of  Eq.(\ref{GSst}) under the scaling symmetry $X_1$ (\ref{sy1}), an O.D.E., which is given by
\begin{equation}\label{Dgen1} 
(8 y^3 - 12 y^2  + 4 y)\, \phi^{\prime \prime}   + (12 y^2   -10 y)\, \phi^{\prime }   + (- 3 y/2 +1/4 ) \, \phi
 =  \end{equation} 
 $$= 1/{ (4\, {\phi}^3 )} +  {a_2\,  y }/{ {\phi}^7  } .  $$
For $  a_2= -3/2$,  Eq.(\ref{Dgen1}) is solved by $\phi \equiv 1$  and returns the result of Sec.\ref{D}.
 \,
 For different values of  $ a_2$, keeping  $\phi(1) = 1$ as a normalization condition
 on $\psi$,  Eq.(\ref{Dgen1})  can be easily solved numerically.  Note that the exceptional symmetry (\ref{exc}) depends only on the scaling of $F(\psi) $ and of $G(\psi)$ on $\psi$ and not on the 
 specific values of the coefficients   $a_0 $ and $ a_2$ { (see also Sec.  \ref{SyGGS})}. Then  the transformation  (\ref{solex}) applies also to the  solutions  of  Eqs.(\ref{Dgen0},\ref{Dgen1})   and sends solutions into new solutions %
 of the form
 \beq\lb{Dsh+}\psi(r,z)\=[ r^2-\big(z+\la  (r^2+z^2) \big)^2]^{1/4} \, \,  \phi( r^2/[r^2-(z+\la   (r^2+z^2) )^2]), \eeq
 which can be rewritten  in a geometrically more transparent way as 
  \beq\lb{Dsh+1}\psi(r,z)/r^{1/2}\= \,  \phi( y_\lambda, a_2)/  y^{1/4}_\lambda,\eeq
  where we have made explicit the dependence of the solution on the parameter $a_2$,  and  \beq  \lb{Dsh+2} y_\lambda \equiv   r^2/[r^2-\big(z+\la   (r^2+z^2) \big)^2], \qquad  1/y_\lambda = 1 -(\tan{\theta} +
   \lambda R/\cos{\theta})^2 ,\eeq
with $R^2 = r^2+z^2$.
\begin{figure}
{\includegraphics[width=.4\textwidth]{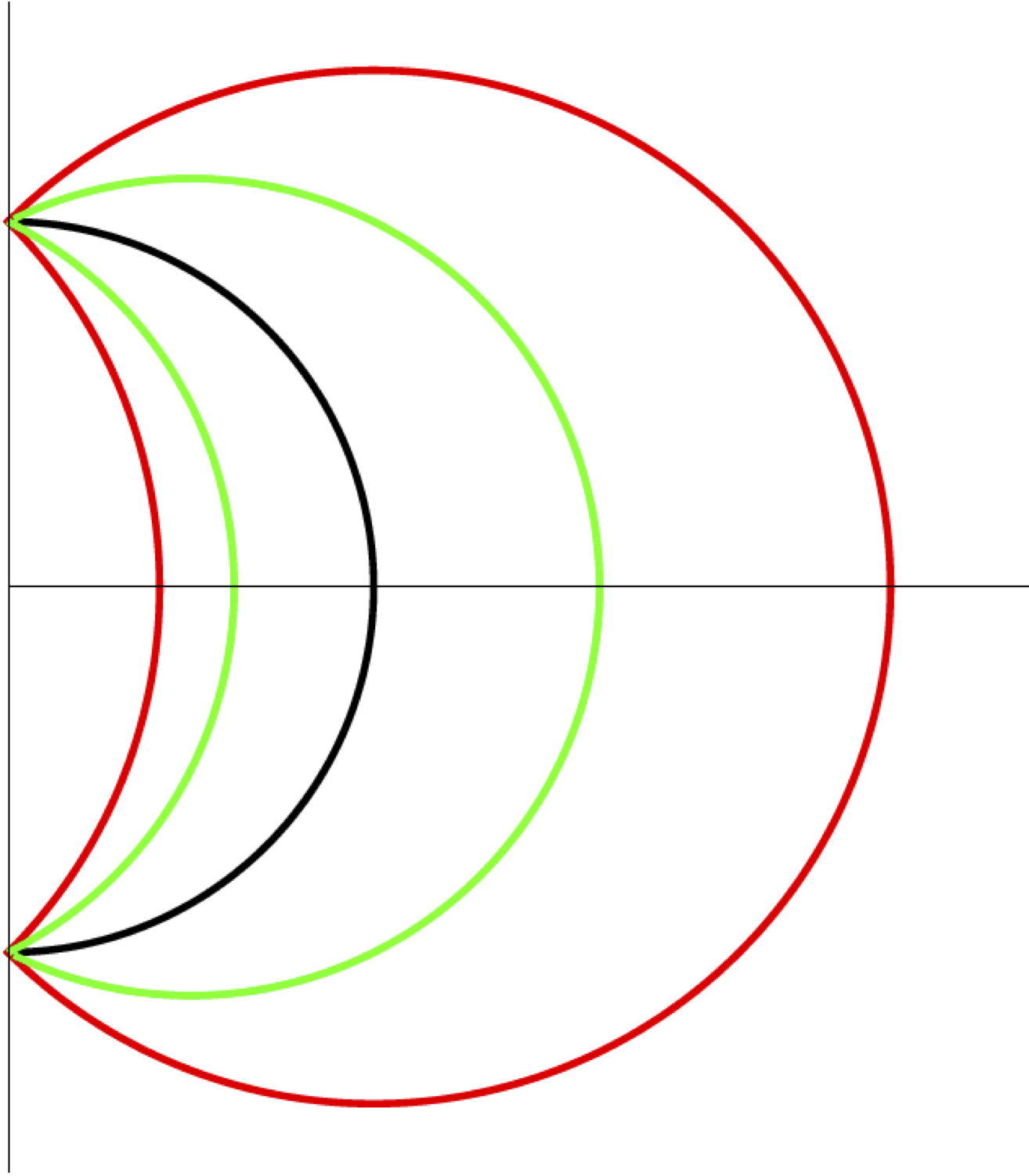}} 
\put(-10,90){$r$}
\put(-200,190){$z$}
		\caption{\small{Lunes with borders given by the circles  in Eq.(\ref{Dsh+3}) for $\lambda =1$ and $1-1/y_\lambda = 1$, red,  $1-1/y_\lambda = 1/2$,  green and  $1-1/y_\lambda = 0$, black. The red border  is the transformed of the lines  $|z| = r$,  the black one  of $z=0$. The  two (common)  vertices of the lunes are the transformed of the origin and of the point at $\infty$.
		}}
\label{lune}
\end{figure}
Notice that  $y=  const.$ corresponds  in the $r\geq 0$  half-plane  to the lines $r= k|z|$, with $k>1$, whereas \,
 $y_\lambda =  const.$ 
corresponds  to the borders of the {\it lunes} (crescents)     defined by the circles  \beq  \lb{Dsh+3} 
\left(\,  r \, \mp \, (1 - 1/y_\lambda )^{1/2}/(2\lambda)\,  \right)^2 \,  +\, z^2= (2 - 1/y_\lambda )/(2\lambda)^2 \qquad {\rm  and} \quad r\geq 0,\eeq
as shown in Fig.\ref{lune} 
(here the variable $z$ has been translated, see below Eq.(\ref{Dsh})).  \, Note that the shape of these lunes does not depend on the value of $a_2$  and that the function $\phi( y_\lambda, a_2)/  y^{1/4}_\lambda$
is constant on their borders. Thus the shape in the $r$-$z$ plane of the class of solutions  discussed below  can be visualized as given by the level curves of $r^{1/2}$ times a function that depends on $a_2$ and that is constant on the borders of the lunes  in Fig.\ref{lune}   whose shape depends only on  the transformation parameter $\lambda$.

Noting that the coefficient of $\phi^{\prime \prime}$ in Eq.(\ref{Dgen1})  vanishes at $y=1$ and restricting to solutions
 where $\phi^{\prime \prime}$  does not diverge,   keeping as mentioned above  $\phi(1) =1$,  we find that the value of  $\phi^{\prime}$ at $y=1$  is given by 
\beq\lb{init}  2 \, \phi^{\prime}(1) = 3/2  + a_2 .  \eeq
Two classes of solutions  of Eq.(\ref{Dgen1}) can be identified depending on the value of $a_2$.\,\,
For $a_2  > - 3/2$, and then $\phi'(1)>0$,  the function $\phi$ remains positive  over the whole interval $1 \leq y  \leq  \infty$ and behaves asymptotically as $\phi (y) \propto y^{1/4}$ for $y\to \infty$
 i.e., for $z^2 \to r^2$ . Note that for $a_2=-1$, one has the exact solution  $\phi =  y^{1/4}$  of Eq.(\ref{Dgen1}),  which yields  the special solution  $\psi (r,z)=  r^{1/2}$.   \\For $a_2  < - 3/2$, the function $\phi (y) $  reaches  zero at  a finite value $ {\bar y} =  {\bar y}  (a_2) $.   
For $y\to{\bar y} $, 
 i.e., for $z^2 \to r^2 ( 1 - 1/{\bar y}) $, it  behaves locally  as $\phi (y)  \propto  ( {\bar y} -y)^{1/4}$ which corresponds to  
 $\psi (r,z)\propto [z^2 - r^2 (1- 1/{\bar y})]^{1/4}$.  \,  In this case a complementary solution  can be found for $  {\bar y} < y < \infty $  which starts  as $\phi (y)  \propto  (y-  {\bar y} )^{1/4}$
and behaves as $\propto  y ^{1/4}$ for $y\to \infty$.\\
The structure in the $r$-$z$ half-plane of the  solutions  with  $a_2  > - 3/2$,  after being transformed according to Eq.(\ref{solex}), is similar to that of the D-shaped solution obtained in Ref.\onlinecite{PoP10},
with the magnetic axis shifting towards larger values of $r$ with increasing values of  $a_2$. The structure of the  solutions  with  $a_2  < - 3/2$ will be discussed in the next section in the frame of the solutions of the GGS equation.

\bigskip

\section{Equilibria with flows}
\subsection{Symmetries of the GGS equation}\label{SyGGS}

We now consider the case of the generalized GS equation (\ref{ecce}). 
\\
It is easy to show that the two \sys\ $X_1$ and $X_2$ given above (\ref{sy1}) and (\ref{exc}) 
are still valid: more precisely, the \sy\ $X_1$ is admitted by any \eq\  of the form
\beq\lb{ggs1} {\pd^2\chi\ov{\pd r^2}}-{1\ov r}{\pd\chi\ov{\pd r}}
+{\pd^2\chi\ov {\pd z^2}}\=\sum_\ell a_\ell\, r^\ell\chi^{1+\frac{2+\ell}{2q}}\eeq
and, similarly, the exceptional \sy\ $X_2$ is admitted by any \eq\ of the form
\beq\lb{ggs2} {\pd^2\chi\ov{\pd r^2}}-{1\ov r}{\pd\chi\ov{\pd r}}
+{\pd^2\chi\ov {\pd z^2}}\=\sum_\ell a_\ell\, r^\ell\chi^{-3-2\ell}\eeq
where at the r.h.s.  any combination of terms $r^\ell\chi^{1+\frac{2+\ell}{2q}}$ (resp. $r^\ell\chi^{-3-2\ell}$) with
arbitrary constants $a_\ell$   and -- in principle -- any real value of $\ell$ is admitted

Restricting to the cases relevant to Eq.(\ref{ecce}), we can write instead of (\ref{ggs1})
\beq\lb{ggs1a} {\pd^2\chi\ov{\pd r^2}}-{1\ov r}{\pd\chi\ov{\pd r}}
+{\pd^2\chi\ov {\pd z^2}}\=a_0\chi^{1+\frac{1}{q}}+a_2r^2\chi^{1+\frac{2}{ q}}+a_4r^4\chi^{1+\frac{3}{q}}\eeq
and finally, with $q=-1/4$,
\beq\lb{1} {\pd^2\chi\ov{\pd r^2}}-{1\ov r}{\pd\chi\ov{\pd r}}
+{\pd^2\chi\ov {\pd z^2}}\=\frac{a_0}{\chi^{3}}+\frac{a_2r^2}{\chi^{7}}+\frac{a_4r^4}{\chi^{11}}
\ .\eeq

\subsection{Solutions of the GGS equation }

We proceed as in Sec. \ref{AES} by defining  
\begin{equation}\label{ribollita}  {\chi}(r,z) = s ^{1/4}\,   \phi(y), \quad y = r^2/s  \ge 1,   \quad   \phi(1) =1,\end{equation}
and obtain an O.D.E. (thanks again to the invariance of Eq.(\ref{1}) under the symmetry $X_1$) which is of the form of Eq.(\ref{Dgen1})  with an additional nonlinear term (as before we put $a_0=1/4$):
\begin{equation}\label{Dgen1bis} 
(8 y^3 - 12 y^2  + 4 y)\, \phi^{\prime \prime}   + (12 y^2   -10 y)\, \phi^{\prime }   + (- 3 y/2 +1/4 ) \, \phi
 =  \end{equation} 
 $$= 1/(4 {\phi}^3 ) +  {a_2\,  y }/{ {\phi}^7  }  +  {a_4 \,y^2}/{ {  \phi}^{11} }.$$
As in Sec.(\ref{AES})   the exceptional symmetry allows us to construct a family of D-shaped equilibria\cite{note1} which  now include plasma flows
\beq\lb{DshF+1}\chi(r,z)/r^{1/2}\= \,  \phi( y_\lambda, a_2,a_4)/  y^{1/4}_\lambda .\eeq
These solutions can again be visualized as  the level curves of $r^{1/2}$ times a function that now depends on both $a_2$ and  $a_4$  and that is constant on the borders of the  lunes  defined in Sec.\ref{AES}.\\
Note that for $a_4 \not= 0$ there is no constant solution $\phi = 1$, while Eq.(\ref{init}) is changed into
\beq\lb{initfl}  2 \, \phi^{\prime}(1) = 3/2  + a_2  +a_4.  \eeq

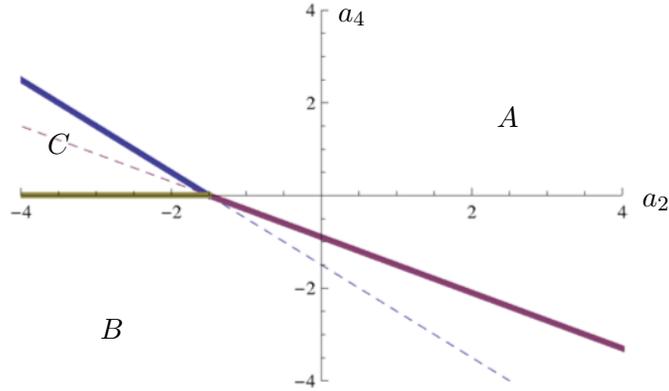
\begin{figure}[h]
\centering
{\includegraphics[width=.5\textwidth]{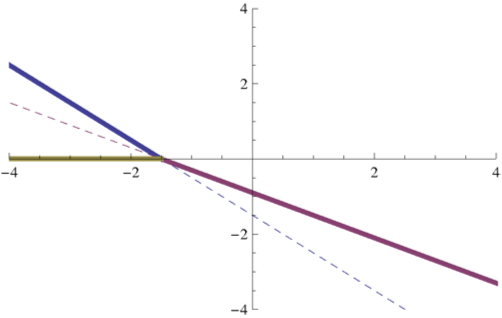}} 
\put(5,70){$\small{a_2}$}
\put(-110,140){$\small{a_4}$}
\put(-50,100){$\huge{A}$}
\put(-220,90){$\huge{C}$}
\put(-200,20){$\huge{B}$}
\caption{\small{Regions  defining the three different types of solutions of Eq.(\ref{Dgen1bis}) in the $a_2$ and $a_4$ parameter plane. For $a_4 =0$ see also Sec.\ref{AES}}}
\label{regions}
\end{figure}

\begin{figure} [!]
\centering
{\includegraphics[width=.6\textwidth]{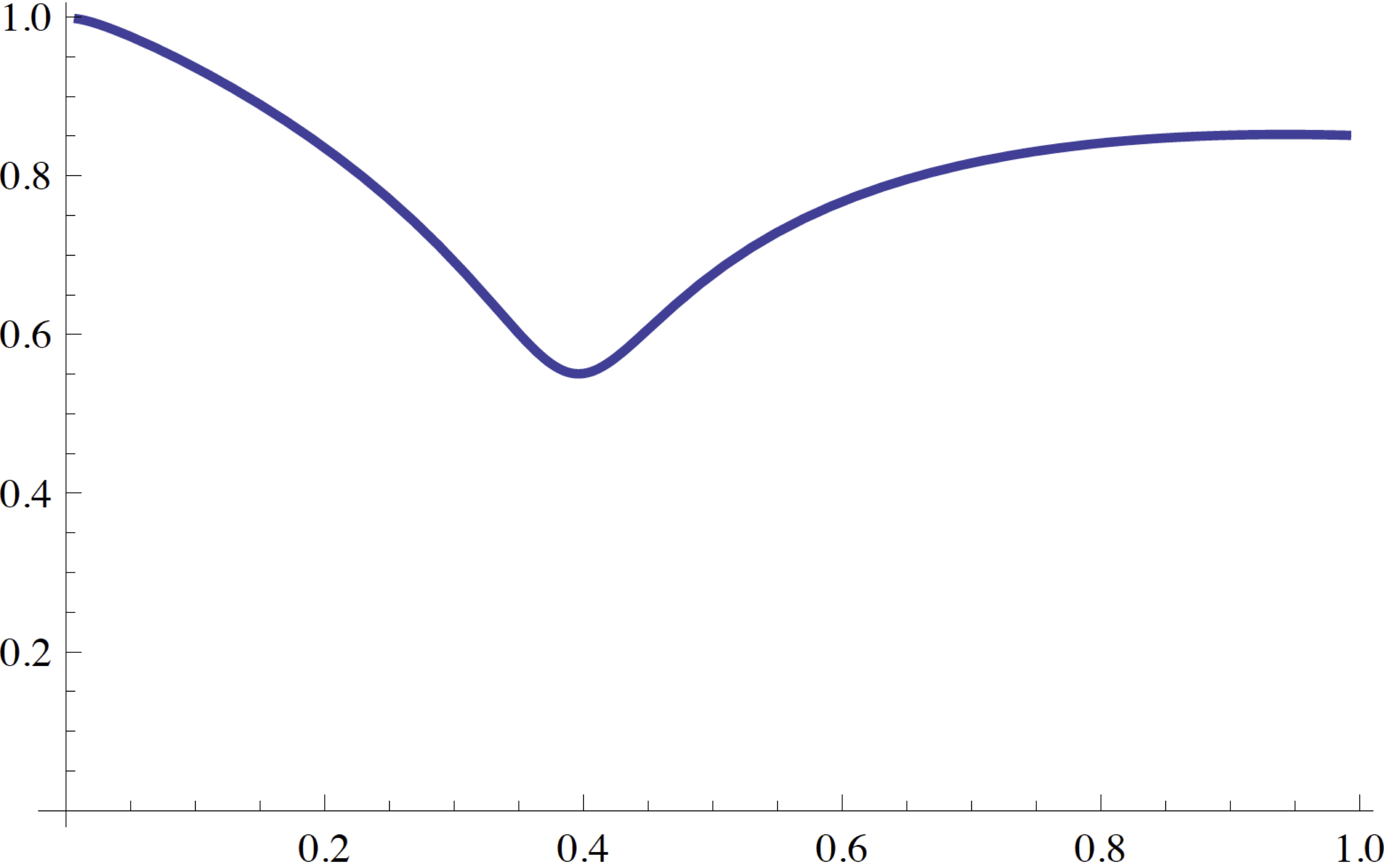}} 
\put(-330,120){{$\phi(y)/y^{1/4}$}}
\put(0,20){$1-1/y$}
\caption{\small{Plot of   $\phi(y,-3.5,0.5) / y^{1/4}$, versus  the variable $1-1/y$.}}\label{rimb}
\end{figure}

 Three classes of solutions  of Eq.(\ref{Dgen1bis}) can be identified depending on the values of $a_2$ and $a_4$  as indicated  in Fig.\ref{regions}.
\\ For $\phi'(1)=(a_4+a_2+3/2)/2\ge 0$ with the additional condition  indicated by numerical calculations $a_4+\rho(a_2+3/2)\gtrsim 0$  with $\rho\simeq 0.85$,  region $A$ of Fig.\ref{regions}, the function $\phi$ remains positive   monotonically growing  over the whole interval $1 \leq y  \leq  \infty$  and behaves asymptotically as 
$\phi (y) \propto y^{1/4}$ for $y\to \infty$ as in the case without flow. 
\begin{figure}[h]
\centering
{\includegraphics[width=.6\textwidth]{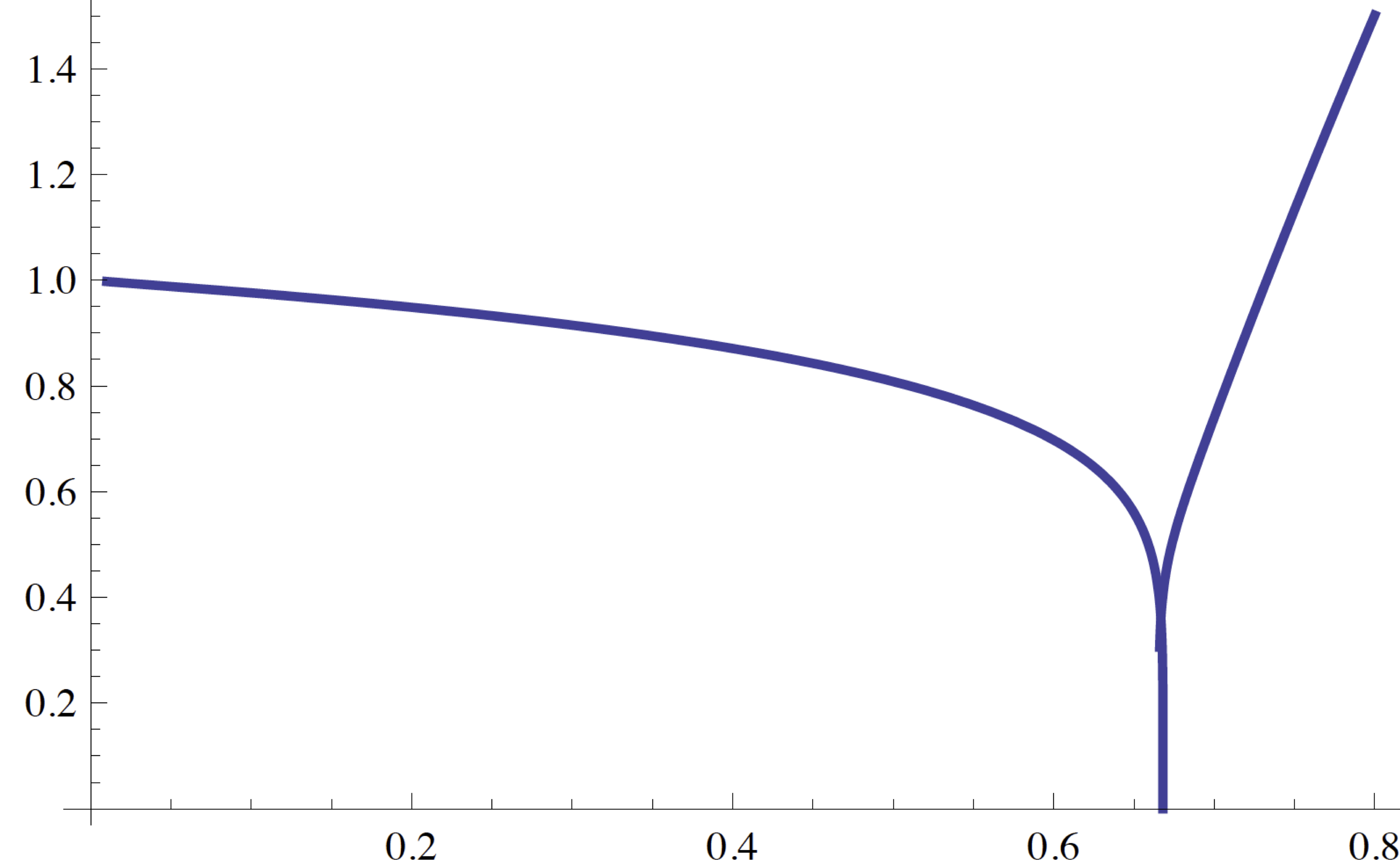}} 
\put(-330,100){{$\phi(y)/y^{1/4}$}}
\put(0,20){$1-1/y$}
\caption{\small{Plot of   $\phi(y,0.5,-1.85) / y^{1/4}$, see Eq.(\ref{DshF+1}),  expressed as a function  of the variable $1-1/y$  whose value determines the  lune borders in Fig.\ref{lune}. Part of the solution for $y > {\bar y}$ is not shown.}}
\label{zerocross}
\end{figure}
\\For 
$a_4+\rho(a_2+3/2)\lesssim 0$ and $a_4<0$, region $B$ of Fig.\ref{regions},  the solution reaches zero at  a finite value 
$y= {\bar y}$, as shown in Fig.\ref{zerocross} in terms of the variable $1-1/y = z^2/r^2$.  
In this case,  $\phi (y)$    behaves locally  as $\phi (y)  \propto  ( {\bar y} -y)^{1/6}$ and, as in the case without flow,   a complementary solution  can be found for $  {\bar y} < y < \infty $  which starts  as $\phi (y)  \propto  (y-  {\bar y} )^{1/6}$ and behaves   for $y\to \infty$ as $\propto  y^{1/4}$. 
\\For $a_4+a_2+3/2< 0$ and $a_4>0$,  region $C$ of Fig.\ref{regions},  we have a solution which remains strictly positive  but does not behave monotonically, and asymptotically grows  as $\propto  y^{1/4}$, as shown in Fig.\ref{rimb} in terms of the variable $1-1/y $. For $a_4+\rho(a_2+3/2)\simeq 0$,
there are  ``separating solutions'' which behave asymptotically as $y^{1/12}$. Notice the particular case $a_4=0,\,a_2=-3/2$ where one recovers the known solution $\phi\equiv  1$.

\begin{figure} [h]
\centering
{\includegraphics[width=.5\textwidth]{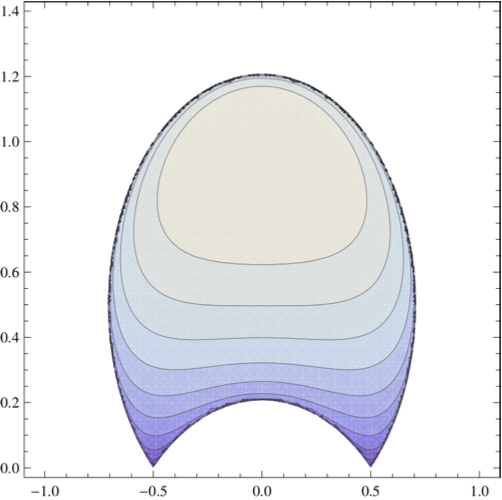}} 
\put(-250,120){{{$r$}}}
\put(-30,20){$z$}
\caption{\small{Shaded contour plot  of the solution $\chi(r,z) \, = \,  r^{1/2}  \phi( y_\lambda,-1.8,0.7)/  y^{1/4}_\lambda$ for $\lambda = 1$.}}\label{1ax}
\end{figure}
In order to illustrate the different shape  in the $r$-$z$ plane of the solutions  belonging to region $A$ and  to regions $B$-$C$, when  transformed according to Eq.(\ref{DshF+1}) and shifted in $z$, we may use simple analytical fitting formulae of the numerical solutions of  $\phi(y,a_2,a_4)$.  
\\The shaded contour plots of two such  solutions  are shown in Figs.\ref{1ax} and \ref{3ax}
 over to whole domain within the red lune  in Fig.\ref{lune}. Obviously, these solutions can be restricted to within a reduced  domain inside this lune by imposing (conducting) boundary conditions on  the   $\chi = const. $ ($\rightarrow \psi  = const.$) border of the chosen domain.
\\ 
The solution from region $A$ with $a_2 =-1.8$,\,  $a_4= 0.7$ and  $\lambda = 1$ is shown in Fig.\ref{1ax}: it  is D-shaped with its magnetic axis shifted toward large $r$  and steep  gradients of $\chi$ near the border of its  outermost magnetic surface.
\\The  solution from region $C$ with $a_2=-3.5$,\, $a_4 =0.5$ (the same of Fig.\ref{rimb}) and  $\lambda = 1$ is shown in Fig.\ref{3ax}.  It exhibits  three  magnetic axes along  the toroidal direction. Thus this solution     corresponds  to a configuration made of  three  sets of surfaces, each set nested around one of the magnetic axes. Each set  can be considered as an independent equilibrium configuration.  The configuration centered at  the middle axis is  D shaped while the outer configuration  has a crescent shape.  A similar structure,  but with  steeper gradients,  is obtained for  solutions from region $B$. In this latter case the three sets of surfaces are separated by the $\chi = 0$ surface.
\begin{figure} [h]
\centering
{\includegraphics[width=.5\textwidth]{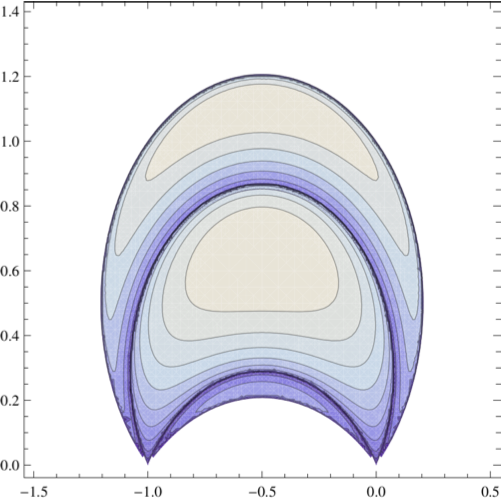}} \label{3ax}
\put(-250,120){{{$r$}}}
\put(-30,20){$z$}
\caption{\small{Shaded contour plot  of the solution $\chi(r,z) \, = \,  r^{1/2}  \phi( y_\lambda,-3.5,0.5)/  y^{1/4}_\lambda$ for $\lambda = 1$.}}\label{3ax}
\end{figure}
\section{Conclusions} 
 
 In this article we have developed a method based on the Lie symmetries of  the GS and of the GGS equations that  we have identified by imposing specific relationships  between  the flux functions  of the magnetic flux variable $\psi$  in these equations.
 
 In the case of the GS equation (plasma equilibria without flows) we have extended the class of the D-shaped solutions found in Ref.\onlinecite{PoP10}. Given the numerical solution of an ODE whose general properties have been discussed in Sec.\ref{AES}, these solutions can be worked out explicitly  in order to obtain the dependence of the plasma pressure and of the toroidal field on the cylindrical coordinate $r$ and $z$ as done  in Ref.\onlinecite{PoP10}.

 In the case of the GGS equation, when treating equilibria with poloidal flows,  we have made use of an implicit integral  transformation  from $\psi$ to a new dependent variable $\chi(\psi)$ that can be applied when the  alfv\`enic Mach number of the poloidal  flow is a flux function and is smaller than unity.  Although the shape of $\chi$ (and thus of $\psi$)  can be  found with  a relatively minor generalization of the procedure adopted for the GS equation, finding the  explicit dependence of the plasma pressure,   of the poloidal and toroidal velocity  fields and of the toroidal field  on $r$ and $z$  in the GGS  case  requires the solution of a  set of  nonlinear equations and,  in the case of poloidal flows, the inversion of the integral relationship between $\phi$ and $\chi$  once the flux function ${\cal F} (\psi)$ is chosen. The system of  Eqs.(\ref{ecce1},\ref{ecce2},\ref{ecce3}) must in fact be solved for the remaining flux functions  to make the functions ${\cal A}_0, {\cal A}_2, {\cal A}_4$ compatible  with Eq.(\ref{ggs1a}). 
 \begin{figure}[h]
{\includegraphics[width=.4\textwidth]{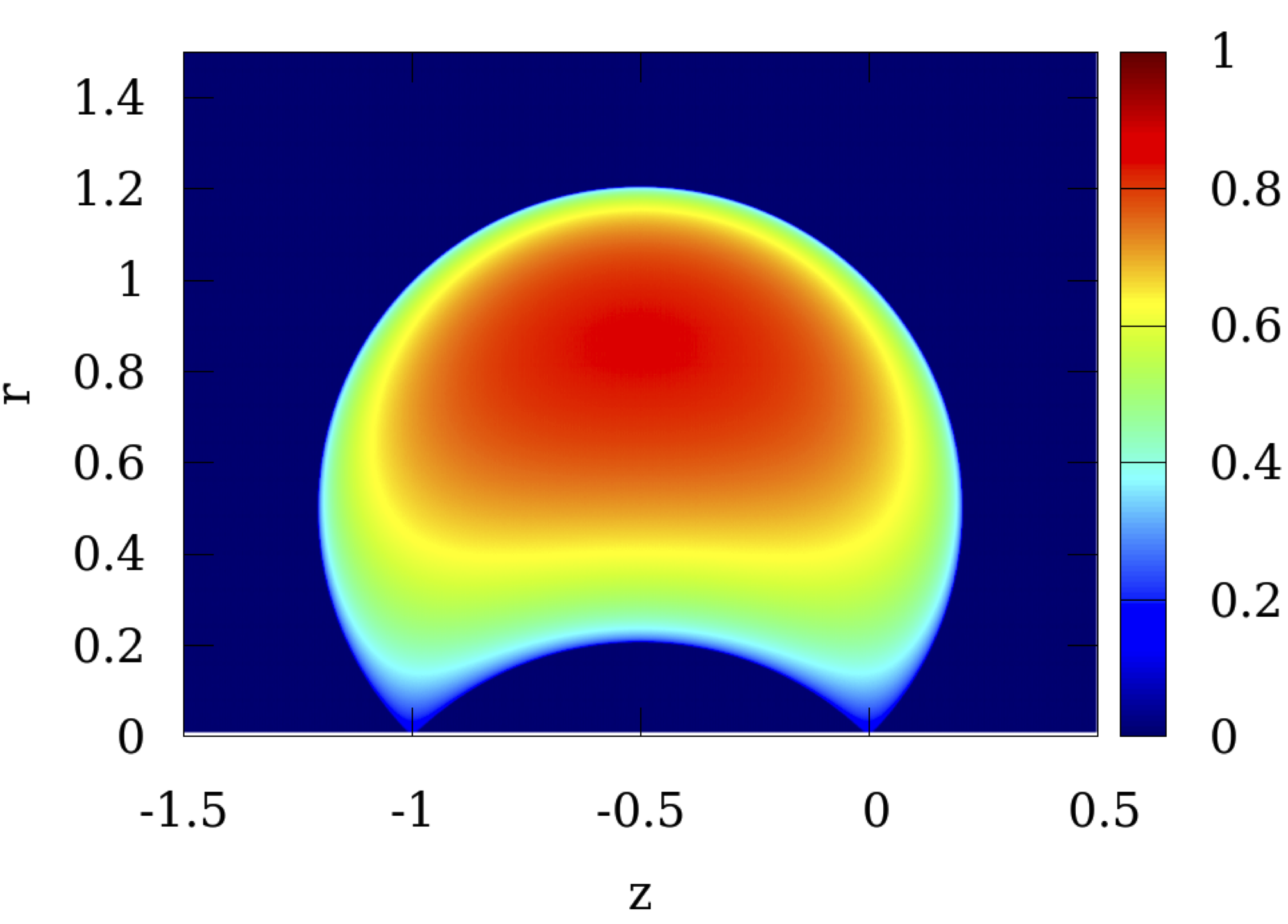}} \\
{\includegraphics[width=.4\textwidth]{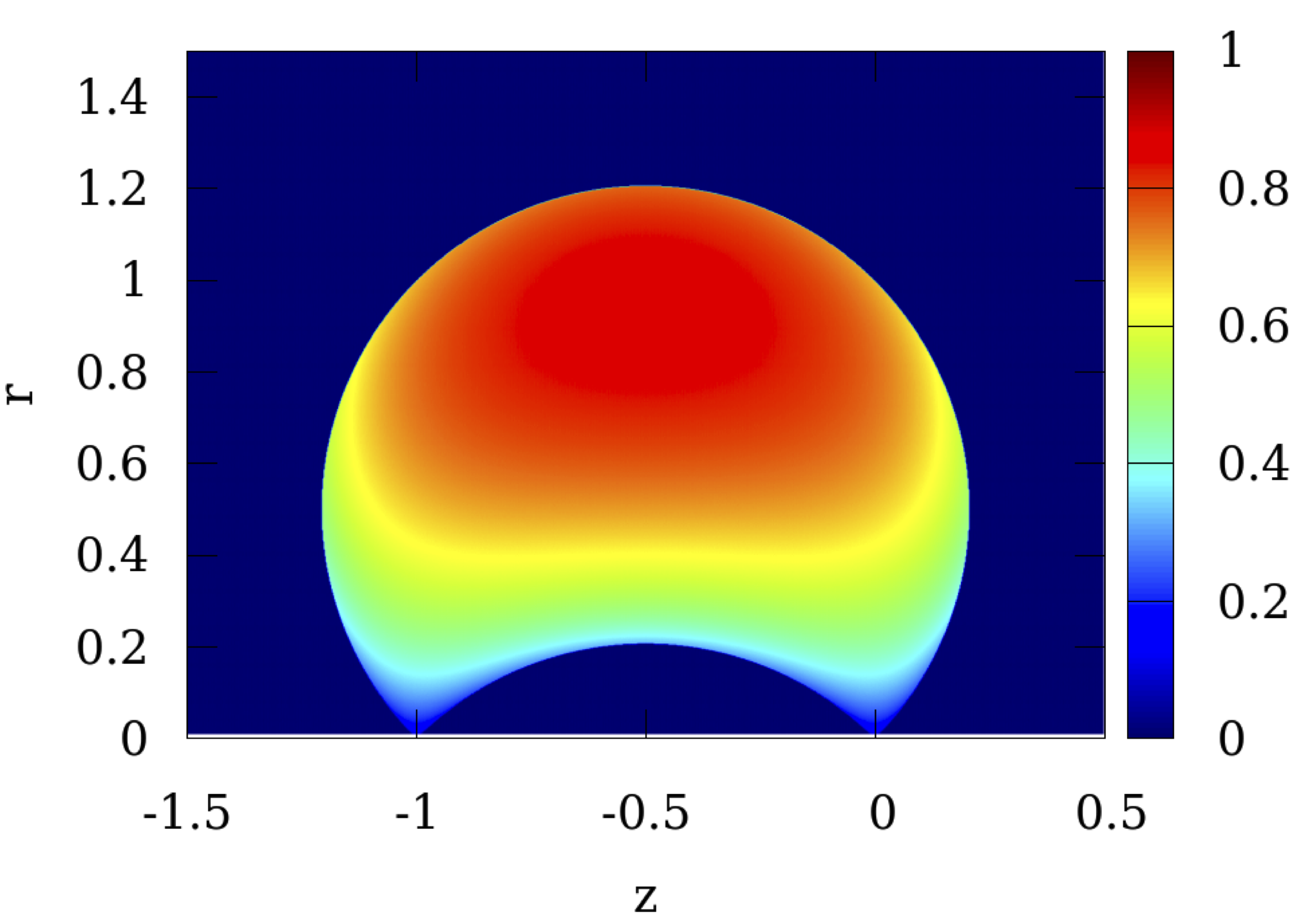}} \\
{\includegraphics[width=.4\textwidth]{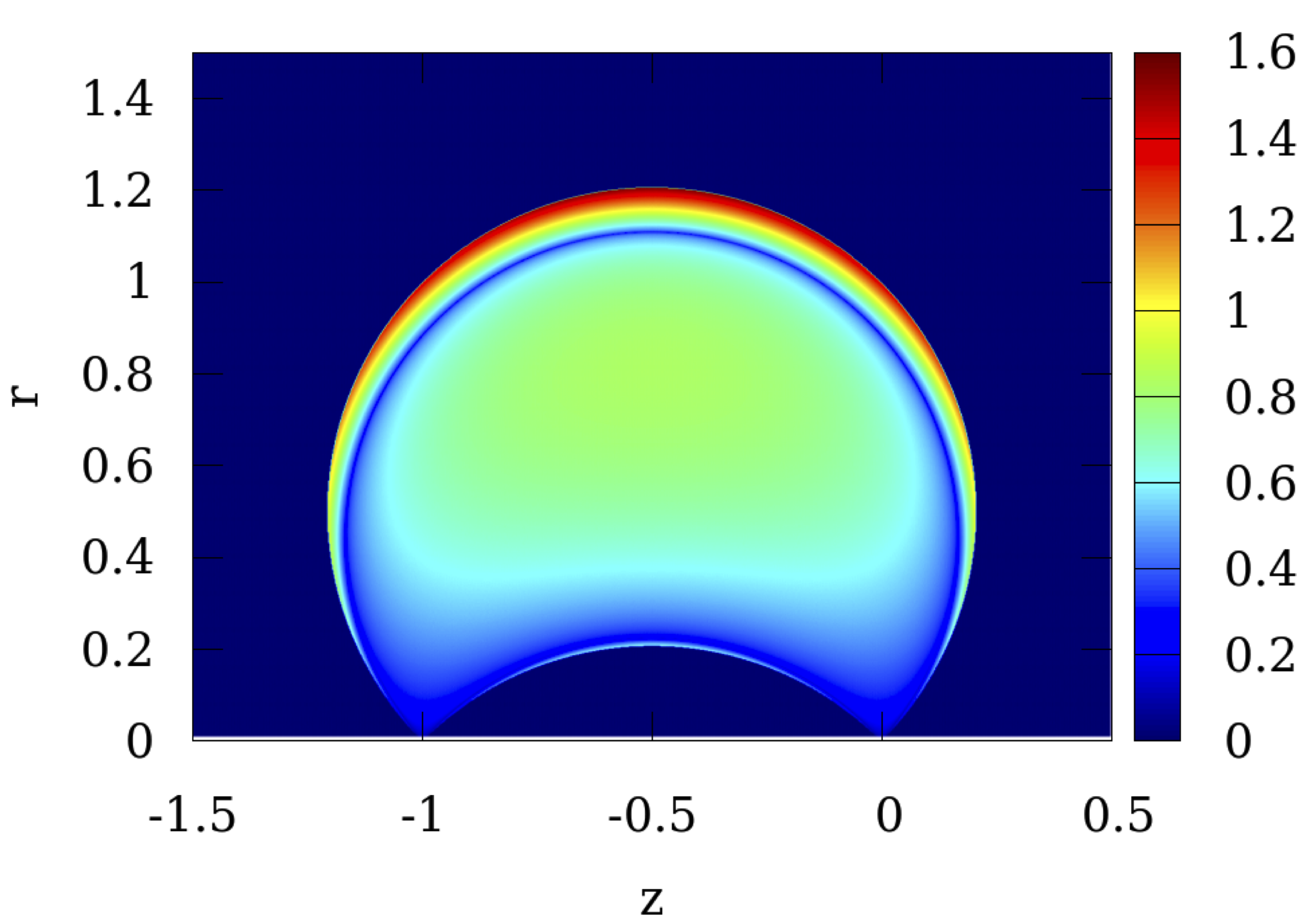}}
\caption{\small{Shaded contourplots of $\psi (r,z)$ for $\lambda = 1$,\, $a_2 = -3/2$ and $a_4 = 0$ top frame,  $ a_4 =  0.1 $ center frame and  $ a_4 = -0.1 $  bottom frame.  }}
\label{compar}
\end{figure} 
\\ This aspect  of the problem, together with  possible restrictions it might bring on the physical range over which the  coefficients $a_2$ and $a_4$ can vary,  has not been addressed  explicitly in the present  article. However  if  physical boundary conditions are imposed,  for example by external  conductors,  that restrict the solution  to  the ``flat'' central domain of e.g.,  the configuration  shown in Fig.\ref{1ax},  that is to say  to a domain inside   $\chi = const $ (and thus $\psi = const$) surfaces within which  $\chi$ is sufficiently larger than zero and the Mach number sufficiently smaller than one,   Eq.(\ref{Morr}) can be inverted by a simple perturbation procedure.
On the contrary, the inverse  power dependence on $\chi$ of the terms of the  terms on the r.h.s. of GGS equation  (\ref{1})  implies that the equilibrium quantities,   including the flow profiles,  will develop  large gradients,  and/or the Mach number will approach unity,  at  the borders of the lunes in Fig.\ref{lune}  where $\chi \to 0$ (and, for solutions in region B, near the internal   $\chi = 0$ curves). This  can be seen by  combining  $d/ d \psi = [ 1 - M^2(\psi)]^{1/2} \, d/d \chi$  with the $\chi $ derivatives of the flux functions ${\cal A}_0 (\chi), {\cal A}_2 (\chi) $ and ${\cal A}_4 (\chi)$ in  Eqs.(\ref{ecce1}-\ref{ecce3}).

Furthermore, it is instructive to   compare the D shaped equilibrium configuration without flows discussed in Ref.\onlinecite{PoP10} with two corresponding equilibria (at fixed ${\cal  J}(\psi)$, see Eq.(\ref{eq:ffun5})) with toroidal flows but without poloidal flows so that the functions $\chi$ and $\psi$ coincide. This comparison is shown in Fig.\ref{compar} where\cite{note2}  the shaded contour plot of the D-shaped equilibrium  of  Ref.\onlinecite{PoP10} is shown in the top frame  ($a_2 = -3/2$, $a_4=0$), that of a corresponding  equilibrium  with angular velocity increasing with  $\psi$  ($a_2 = -3/2$, $a_4> 0 $) is shown  in the center frame  and one with angular velocity decreasing with $\psi$  ($a_2 = -3/2$, $a_4<0 $) in the bottom frame.
  \\ We see that a positive  gradient of the centrifugal term  due to the toroidal flow ($a_4 >0$)  leads to a shift of  the maximum of the flux function $\psi(r,z)$  towards larger values of $r$ and  to the formation of   large gradients at the outer border, as characteristic of the equilibrium solutions  in region A (see e.g., Fig.\ref{1ax}). On the contrary, a negative  gradient of the centrifugal term  ($a_4 <0$) causes $\psi(r,z)$ to vanish not only at the boundary of the solution domain but also along two  curves inside the domain as is the case for the solutions in region C. This splits  the solution into  ``separate''  solutions with different magnetic axes, as characteristic of the equilibrium solutions  in regions B and  C (see e.g.,  Fig.\ref{3ax}). From the bottom frame of Fig.\ref{compar} we see that the inner solution is,  for the chosen parameters, vanishingly thin, the central one  has a rather flat radial profile  while the outer solution has large gradients. 

As a concluding  remark we observe that  the  procedure  developed in the present paper   can be further  extended to a wider class of solutions by adapting it  to the  different choices of the variable $s= s(r,z)$ made  in Sec.IV of Ref.\onlinecite{PoP10}.

\section*{Acknowledgments}

We thank G.N. Throumoulopoulos for pressing us into extending the results we presented in Ref.\onlinecite{PoP10} to the case of equilibria with flows and T. Andreussi for clarifying discussions on the structure   of the flux functions in the GGS equation.


\begin{thebibliography}{99}
\bibitem{ext} An extension to super-alfv\`enic flows is treated in Ref.\onlinecite{GK}, including cases where the critical point is crossed.
\bibitem{note} {The ``complementary'' solution $\psi = (z^2-r^2)^{1/4}$, see Ref.\onlinecite{PoP10},  could have  also  been considered here and in the developments of the following sections. For the sake of brevity this case is not
discussed explicitly in the present article.}
\bibitem{note1} {Since $\psi = \psi(\chi)$, as follows from inverting Eq.(\ref{Morr}), the shape of the $\psi$ surfaces is the same as that of the $\chi$ surfaces.}
\bibitem{note2} In Fig.\ref{compar} the exact numerical solutions of Eq.(\ref{Dgen1})  are used while in Figs.\ref{1ax} and \ref{3ax} simple analytical fitting formulae    are used.
\end{thebibliography}
\end{document}